\documentclass[aps,prx,reprint,superscriptaddress,amsmath,amssymb,showpacs,floatfix,longbibliography]{revtex4-1}
\usepackage{epsf,amsmath,amssymb,amsfonts,verbatim,color,multirow,pifont}
\usepackage{graphicx}
\usepackage{bm}
\usepackage{dcolumn}
\usepackage{txfonts}
\usepackage{hyperref}
\usepackage{soul}

\begin{document}
\title{Universal form of thermodynamic uncertainty relation for Langevin dynamics} 
\author{Jae Sung Lee}
\email{jslee@kias.re.kr}
\author{Jong-Min Park}
\author{Hyunggyu~Park}
\email{hgpark@kias.re.kr}
\affiliation{School of Physics and Quantum Universe Center, Korea Institute for Advanced Study, Seoul 02455, Korea}
\newcommand{\revise}[1]{{\color{red}#1}}

\date{\today}

\begin{abstract}
Thermodynamic uncertainty relation (TUR) provides a stricter bound for entropy production (EP) than that of the thermodynamic second law. This stricter bound can be utilized to infer the EP and derive other trade-off relations. Though the validity of the TUR has been verified in various stochastic systems, its application to general Langevin dynamics has not been successful in a unified way, especially for underdamped Langevin dynamics, where odd parity variables in time-reversal operation such as velocity get involved. Previous TURs for underdamped Langevin dynamics is neither experimentally accessible nor reduced to the original form of the overdamped Langevin dynamics in the zero-mass limit. Here, we find an operationally accessible TUR for underdamped Langevin dynamics with an arbitrary time-dependent protocol. We show that the original TUR is a consequence of our underdamped TUR in the zero-mass limit. This indicates that the TUR formulation presented here can be regarded as the universal form of the TUR for general Langevin dynamics. The validity of our result is examined and confirmed for three prototypical underdamped Langevin systems and their zero-mass limits; free diffusion dynamics, charged Brownian particle in a magnetic field, and molecular refrigerator.
\end{abstract}

\pacs{05.70.-a, 05.40.-a, 05.70.Ln, 02.50.-r}

\maketitle

\section{Introduction}
Thermodynamic processes and accompanying entropy production (EP) are constrained by the thermodyanmic second law, stating that the EP  is always nonnegative.
Beyond the second law, a new thermodynamic bound was discovered in 2015~\cite{Barato}, called the thermodynamic uncertainty relation (TUR) expressed in terms of the TUR factor ${\cal Q}$ as
\begin{align}
	\mathcal{Q}\equiv \frac{\textrm{Var}[\Theta]}{\langle \Theta \rangle^2} \Delta S^\textrm{tot} \geq 2 k_\textrm{B}, \label{eq:originalTUR}
\end{align}
with a time-accumulated current $\Theta$, its steady-state average $\langle \Theta\rangle$ and variance $\textrm{Var}[\Theta]$ , the Boltzmann constant $k_\textrm{B}$, and the average total EP $\Delta S^\textrm{tot}$.
This is basically a trade-off relation between the fluctuation magnitude and the thermodynamic cost of a stochastic system given as an inequality with the universal lower bound. As the variance is always positive, the TUR sets a positive lower bound of the EP, thus provides a tighter bound than the second law. This bound can be utilized for inferring the EP by measuring a certain current statistics in a nonequilibrium process~\cite{Fakhri, Manikndan, Gingrich2}. Moreover, a recent debate on thermodynamic trade-off relations among the efficiency, power, and reversibility of a heat engine~\cite{Benenti, Brandner, Karel, Campisi, Shiraish, Holubec1,JSLee,JSLee1} has also been investigated based on the TUR bound~\cite{Pietzonka}.

After the first discovery in 2015~\cite{Barato}, the validity of the TUR has been rigorously proven for a variety of stochastic systems~\cite{Gingrich1, Horowitz, Hasegawa2019, dechant2018, Ueda, Seifert2020, Proesmans, Samuelsson, hysteretic, Barato2, FPS, FT_TUR,Macieszczak}. First, it was shown that the TUR in the original form, Eq.~\eqref{eq:originalTUR}, holds for a continuous-time Markov process with discrete states~\cite{Gingrich1, Horowitz} and the overdamped Langevin dynamics with a continuous state space in the steady state~\cite{Hasegawa2019}. Later, TURs for these two stochastic systems with an arbitrary initial state~\cite{dechant2018, Ueda} and an arbitrary time-dependent driving~\cite{Seifert2020}  have been found. The TUR for a discrete-time Markov process was first discovered only in an exponential form~\cite{Proesmans}, but later, the linearized version was also found~\cite{Ueda}.
We note that the TUR for general stochastic systems was found in an exponential form recently~\cite{Samuelsson, hysteretic}. However, the exponential form is not practically useful in a sense that the physical meaning of the cost function is hard to be interpreted and its bound is quite loose far from equilibrium due to the nature of the exponential function.

Compared to other stochastic systems, studies on the TUR for underdamped Langevin systems have made little progress. In contrast to the overdamped Langevin systems, the odd-parity variables like velocity come into play in the underdamped dynamics
and the probability current is divided into two parts; the reversible and the irreversible current. As only the latter contributes to the EP~\cite{risken1989fpe, Sasa_Entropic}, the thermodynamic cost function could not be simply written in terms of the EP only, but also includes some kinetic quantities such as dynamical activity, which are not easily accessible in experiments~\cite{Hasegawa_underdamped, JSLee_underdamped}. This significantly degrades the applicability of the TUR for inferring the EP in the underdamped Langevin dynamics.
In addition, the link between the TURs for the overdamped and underdamped Langevin dynamics has been missing. Mathematically, the overdamped dynamics is usually attained in the zero-mass limit of the underdamped dynamics. However, the zero-mass limit of the previous TURs for the underdamped dynamics becomes meaningless as the dynamic activity (thus, the cost function) diverges~\cite{JSLee_underdamped}. This clearly reveals the lack of systematic understanding on the thermodynamic trade-off relation in a more fundamental level of description. Moreover, due to this difficulty, the TUR for the underdamped Langevin dynamics with an arbitrary time-dependent driving force has not been studied.

In this study, we derive rigorously an operationally accessible TUR for general underdamped Langevin systems with an arbitrary time-dependent driving protocol, including velocity-dependent forces like a magnetic Lorentz force breaking time reversal.
The cost function of this TUR is expressed in terms of the EP without any kinetic quantity and an initial-state-dependent term
which is negligible for the long observation-time limit. Furthermore, this TUR returns back to the original TUR of the overdamped dynamics (Eq.~\eqref{eq:originalTUR}) in the zero-mass limit when the driving forces and the current weight function do not include odd variables.
Thus, our TUR can be regarded as the universal form of the TUR for general Langevin dynamics.

\section{Model and main results}
We consider a $N$-dimensional underdamped Langevin system driven by a force $\textit{\textbf{F}}(\textit{\textbf{x}},\textit{\textbf{v}},t) = (F_1, \cdots, F_N)$, where $\textit{\textbf{x}} = (x_1, \cdots, x_N)$ and $\textit{\textbf{v}} = (v_1, \cdots, v_N)$ are the position and velocity vectors of the system, respectively. Dynamics of the $i$-th component of the system $(x_i, v_i)$ is in contact with a thermal reservoir with temperature $T_i$. Then, the dynamics can be described by the following equation:
\begin{align}
	\dot{x}_i=v_i,~~m_i \dot{v}_i = F_i (\textit{\textbf{x}},\textit{\textbf{v}},t) - \gamma_i v_i +\xi_i, \label{eq:underdamped_Langevin}
\end{align}
where $m_i$, $\gamma_i$, and $\xi_i$ are the $i$-th mass, dissipation coefficient, and Gaussian white noise satisfying $\langle \xi_i (t) \xi_j (t^\prime) \rangle = 2 k_\textrm{B} \gamma_i T_i \delta_{ij} \delta(t-t^\prime) $ with zero mean, respectively. For convenience, we set the Boltzmann constant $k_\textrm{B} =1$ in the following discussion. A general time-dependent force $F_i$ consists of two parts; reversible $F_i^\textrm{rev}$ and irreversible $F_i^\textrm{ir}$ forces, that is,
$F_i(\textit{\textbf{x}}, \textit{\textbf{v}},t) = F_i^\textrm{rev}(\textit{\textbf{x}}, \textit{\textbf{v}},t) + F_i^\textrm{ir}(\textit{\textbf{x}}, \textit{\textbf{v}},t)$ with
$F_i^\textrm{rev}(\textit{\textbf{x}}, \textit{\textbf{v}},t) = {F_i^\textrm{rev}}^\dagger (\textit{\textbf{x}}, -\textit{\textbf{v}},t)$ and $F_i^\textrm{ir}(\textit{\textbf{x}}, \textit{\textbf{v}},t) = - {F_i^\textrm{ir}}^\dagger (\textit{\textbf{x}}, -\textit{\textbf{v}},t)$, where the `$\dagger$' operation reverses signs of all odd parameters in the time-reversal process~\cite{Sasa_Entropic,JSLee_underdamped}.
Without loss of generality, we can set
\begin{align}
	&F_i^\textrm{rev}(\textit{\textbf{x}}, \textit{\textbf{v}},t) = s f_i^\textrm{rev}(r\textit{\textbf{x}}, \textit{\textbf{v}},\omega t), \nonumber \\
	&F_i^\textrm{ir}(\textit{\textbf{x}}, \textit{\textbf{v}},t) =  f_i^\textrm{ir}(r\textit{\textbf{x}}, \textit{\textbf{v}},\omega t),
\end{align}
where $s$, $r$, $\omega$ are the scaling parameters for force, position, and time, respectively. Note that $s$ is multiplied to  the reversible force only, which is one of the key manipulation for deriving the TUR.
We consider $\Gamma = [\textit{\textbf{x}}_t, \textit{\textbf{v}}_t]_{t=0}^{t=\tau}$, which denotes a trajectory of the system from $t=0$ to $t=\tau$, and a $\Gamma$-dependent current $\Theta$ which has the following form:
\begin{align}
	\Theta_\tau ({\bm \Lambda}) \equiv \int_0^\tau dt~ {\bm \Lambda}(\textit{\textbf{x}}_t,\textit{\textbf{v}}_t,t;s,r,\omega) \cdot \textit{\textbf{v}}_t,
\end{align}
with the weight function vector
\begin{align}
{\bm \Lambda} = s {\bm \chi} (r\textit{\textbf{x}}_t,\textit{\textbf{v}}_t,\omega t)~.
\end{align}
Note that the same scale parameter $s$ is used for the weight function and the reversible force for later convenience.

Then, our first main result is the following {\em underdamped} TUR in terms of  the
underdamped TUR factor $\mathcal{Q}^\textrm{u}$  as
\begin{align}
	\mathcal{Q}^\textrm{u} \equiv \frac{\textrm{Var}[\Theta_\tau]}{\Omega_\tau^2} \left( \Delta S_\tau^\textrm{tot}  + \mathcal{I} \right) \geq 2,\label{eq:main_result}
\end{align}
where $\Omega_\tau$ is defined as
\begin{align}
	\Omega_\tau \equiv \hat{h}_\tau \langle \Theta_\tau\rangle,  ~\textrm{where }\hat{h}_\tau \equiv   \tau \partial_\tau - s \partial_s - r\partial_r -\omega \partial_\omega~,\label{eq:Xi}
\end{align}
and
$\mathcal{I}$ is an initial-state-dependent term defined in Eq.~\eqref{eq:totEPandY} which
depends on the dynamic details but becomes negligible in the large-$\tau$ limit.
Equation~(\ref{eq:main_result}) holds for processes with arbitrary time-dependent driving from an arbitrary initial state.
This underdamped TUR resembles the overdamped TUR recently found in~\cite{Koyuk2020Thermodynamic} with additional
scale parameters $s$ and $r$.
Note that $\Omega_\tau$ is experimentally accessible by measuring the response of $\langle\Theta_\tau \rangle$ with respect to a slight change of the observation time $\tau$, the reversible force magnitude $s$, the system scale $r$, and the driving speed $\omega$.
Thus, the EP can be readily inferred from real experiments by measuring a proper current or a set of currents~\cite{Dechant2019multidimensional}. We emphasize that our underdamped TUR does not contain any kinetic term like dynamical activity.
Furthermore, this TUR provides a much tighter bound, compared to the previous
TURs for the underdamped dynamics~\cite{Samuelsson, hysteretic, Hasegawa_underdamped, JSLee_underdamped}, which will be explicitly shown in the examples below.

Another fascinating part of our undermdaped TUR is that the overdamped TUR, Eq.~\eqref{eq:originalTUR}, arises naturally by taking the zero-mass limit,
in case of no velocity-dependent force.
For simplicity, we consider a steady-state TUR without any time-dependent protocol and no time-dependence in the weight function ${\bm \Lambda}$ of a current of interest ($\omega=0$). To obtain the standard overdamped limit, the velocity variables should not be included in the driving force, i.e.~
\begin{align}
	F_i = s f_i^\textrm{rev} (r \textit{\textbf{x}}) ~\textrm{ and }~ {\bm \Lambda} = s {\bm \chi} (r \textit{\textbf{x}}_t).  \label{eq:overdamped_f_lambda}
\end{align}
Then, in the zero mass limit, $\Omega_\tau$ and $\mathcal{I}$ in Eq.~\eqref{eq:main_result} becomes
\begin{align}
	\Omega_\tau = - \langle \Theta_\tau   \rangle ~\textrm{ and }~ \mathcal{I} = 0 \label{eq:overdamped_Omega}
\end{align}
in the steady state, which leads to the original TUR (Eq.~\eqref{eq:originalTUR}). This is our second main result.
The overdamped TUR for an arbitrary time-dependent protocol is discussed in Sect. \ref{sec:deriation}.
The proofs of Eq.~\eqref{eq:main_result} and Eq.~\eqref{eq:overdamped_Omega} are presented in Sect. \ref{sec:deriation}.

\section{Examples}
To illustrate the usefulness and validity of our main results,
we concentrate on steady-state processes where $\textit{\textbf{F}}$ and ${\bm \Lambda}$ have no explicit time dependence
in the following examples.
With these conditions, the underdamped TUR is simplified with
\begin{align}
	\Omega_\tau=\Omega_\tau ^\textrm{ss} \equiv   (1- s \partial_s - r\partial_r) \langle \Theta_\tau \rangle ~, \label{eq:Xi_ss}
\end{align}
in the steady state.

\subsection{Example 1: free diffusion with drift}
Consider a displacement current in the free diffusion process
of a Brownian particle with mass $m$, driven by a constant external force $\textit{\textbf{F}}$.  We set $\textit{\textbf{F}} = s f \textit{\textbf{e}}_1 = \textit{\textbf{F}}^\textrm{rev}$, where $f$ is a constant and $\textit{\textbf{e}}_1$ is the unit vector along the $x_1$ axis. We choose the weight function ${\bm \Lambda} = s \textit{\textbf{e}}_1$, yielding $\Theta_\tau ({\bm \Lambda})$
as displacement at $t=\tau$ from the initial position at $t=0$ along the $x_1$-axis.
Note that $s$ is a scale parameter, which will be set to be unity after the whole calculation.
This model was studied recently as a paradigmatic example for a conjecture of the underdamped TUR in one dimension~\cite{Chun2}.

It is easy to show that the steady-state velocity is  $\langle v_1 \rangle^\textrm{ss} = sf/\gamma_1$, thus we get	
$\langle \Theta_\tau \rangle  = \tau s\langle v_1 \rangle^\textrm{ss}= \tau s^2  f/\gamma_1$. Consequently, we obtain
\begin{align}
		\Omega_\tau^\textrm{ss} = (1-s \partial_s )\langle \Theta_\tau  \rangle = - \langle \Theta_\tau
\rangle. \label{eq:Xi_ss_free_diff}
\end{align}
Using Eq.~\eqref{eq:main_result} and Eq.~\eqref{eq:Xi_ss_free_diff}, the underdamped TUR for the free diffusion process with drift at $s=1$ becomes

\begin{align}
	\mathcal{Q}^\textrm{u} =\frac{\textrm{Var}[\Theta_\tau]}{\langle \Theta_\tau \rangle^2} \left( \Delta S_\tau^\textrm{tot}  + \mathcal{I}^\textrm{fr} \right) \geq 2,\label{eq:free_diff_TUR}
\end{align}
where $\Delta S_\tau^\textrm{tot}=\tau f^2/(T_1\gamma_1)$ and $\mathcal{I}^\textrm{fr} = 2m  f^2 /(T_1 \gamma_1^2)$. Calculations of $\Delta S_\tau^\textrm{tot}$, $\mathcal{I}^\textrm{fr}$, and $\textrm{Var}[\Theta_\tau]$ are presented in Supplemental Material~\cite{SupplMat}. Note that $\mathcal{I}^\textrm{fr}$ vanishes in the zero-mass limit, confirming that
our underdamped TUR in Eq.~\eqref{eq:free_diff_TUR} returns back to the original TUR form in the overdamped limit. For a finite mass, the original TUR is recovered only when  $\mathcal{I}^\textrm{free}$ is negligible in the large-$\tau$ limit.

\begin{figure}
\includegraphics*[width=\columnwidth]{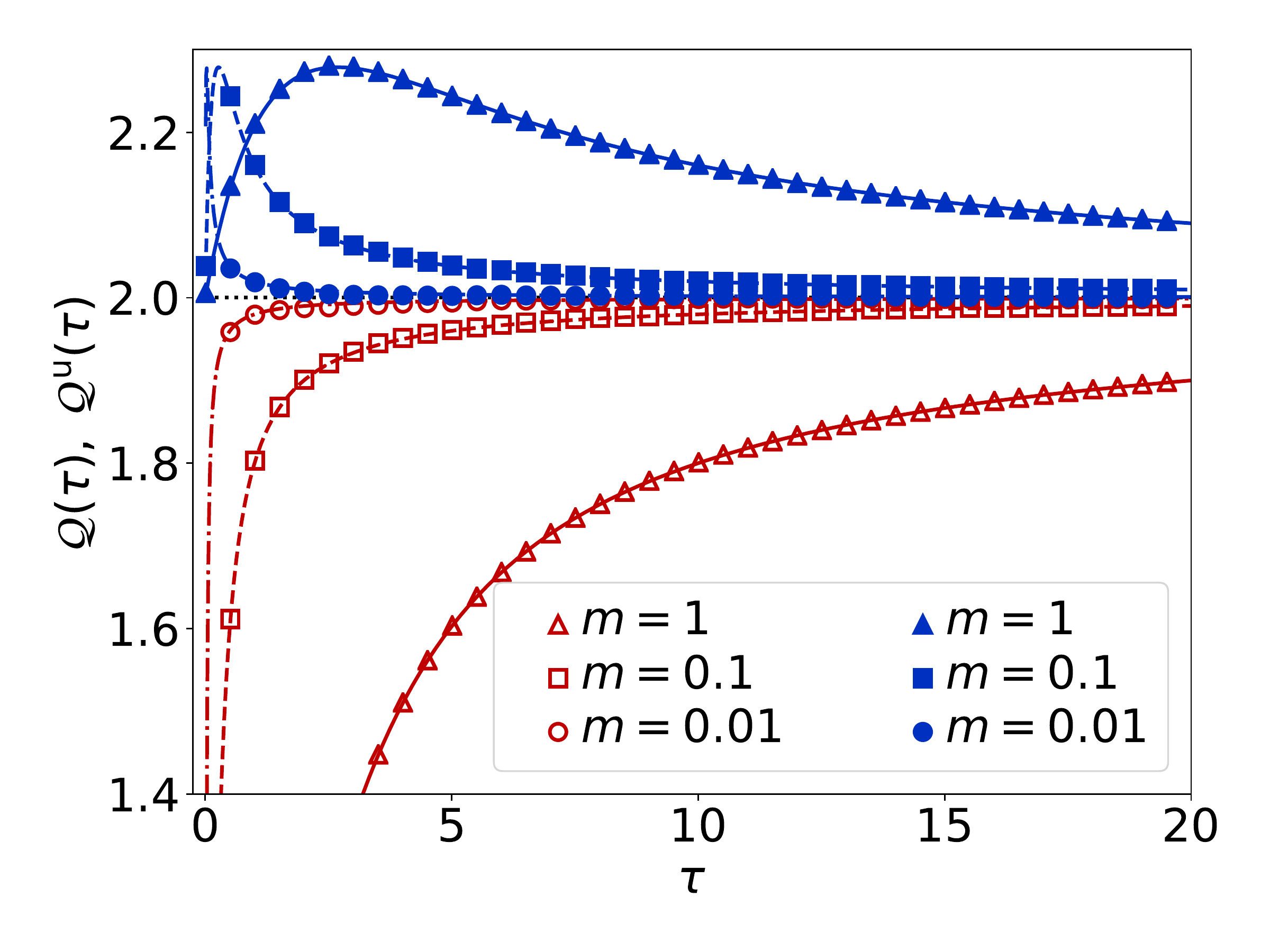}
\caption{ $\mathcal{Q}^\textrm{u}$ (blue filled symbol) and $\mathcal{Q}$ (red open symbol) as a function of the observation time $\tau$ for free diffusion with drift.
Solid, dashed, and dash-dotted curves represent analytic results for $m = 1$, $0.1$, and $0.01$, respectively. The black dotted line indicates the lower bound, i.e.~$2$. Other system parameters are set to be unity, $ f = \gamma_1= T_1 = 1$. }
\label{fig:ex1}
\end{figure}

Figure~\ref{fig:ex1} shows analytic (curves) and numerical (dots) plots of the TUR factors  $\mathcal{Q^\textrm{u}}$ and $\mathcal{Q}$ for various values of $m$ as a function of $\tau$. The analytic expressions are presented in Ref.~\cite{SupplMat} and the numerical data are obtained by averaging over $10^7$ trajectories from the Langevin equation.
As expected from our underdamped TUR, $\mathcal{Q}^\textrm{u}$ is always above the lower bound of 2 for any observation time period $\tau$, and approaches the bound either in the zero-mass limit or in the large-$\tau$ limit.  The conventional TUR factor ${\cal Q}$ approaches the bound from below (violations of the original TUR) in these limits.
This example clearly demonstrates the importance of the initial-state dependent term $\mathcal{I}$ in the underdamped dynamics for a finite $\tau$, which usually vanishes in the overdamped limit. The free-diffusion bound conjecture~\cite{Chun2} also involves $\tau$ in the lower bound, though it differs from our rigorous bound (see Ref.~\cite{SupplMat} for discussions).

\subsection{Example 2: charged particle in a magnetic field}
The next example is the motion of a charged Brownian particle under a magnetic field $B$ in a two-dimensional space~\cite{Chun, JSLee_underdamped}. The particle is trapped in a harmonic potential with stiffness $k$ and driven by a nonconservative rotational force. Then, the total force is given by $\textit{\textbf{F}} = \textit{\textbf{F}}^\textrm{nc} +  \textit{\textbf{F}}^\textrm{mag} + \textit{\textbf{F}}^\textrm{har}$
with  the nonconservative rotational force $\textit{\textbf{F}}^\textrm{nc}= s \kappa (rx_2, -rx_1)$, the Lorentz force induced by the magnetic field $\textit{\textbf{F}}^\textrm{mag} = sB(v_2,-v_1)$, and the harmonic force $ \textit{\textbf{F}}^\textrm{har}= -sk(rx_1,rx_2)$.  By regarding the magnetic field $B$ as an odd-parity parameter, we treat the whole force $\textit{\textbf{F}}$ as a reversible one. The opposite choice is also possible~\cite{Kwon, ChunNoh, JSLee_underdamped}. Here, we consider the case $\gamma_1=\gamma_2  \equiv \gamma$, $m_1 = m_2 \equiv m$, and $T_1 = T_2 \equiv T$. We are interested in the work current done by the nonconservative force, thus, ${\bm \Lambda} = \textit{\textbf{F}}^\textrm{nc}$. By replacing  the parameters as $\kappa \rightarrow s r \kappa$, $B \rightarrow s B$, and $k \rightarrow s r k$ from the result of Ref.~\cite{Chun}, the steady-state work current can be written  as
\begin{align}
	\langle \Theta_\tau \rangle	= \frac{2 \tau r \kappa^2 T}{\gamma k / s + \kappa B - r \kappa^2 m /\gamma}, \label{eq:work_current}
\end{align}
with the stability condition $\gamma k / s + \kappa B - r  \kappa^2 m/\gamma >0$.
Then, we obtain from Eq.~\eqref{eq:Xi_ss}
\begin{align}
		\Omega_\tau^\textrm{ss}
= -\frac{\gamma k /s+r \kappa^2 m/\gamma}{\gamma k /s + \kappa B- r  \kappa^2 m/\gamma} \langle \Theta_\tau\rangle~,
\label{eq:magnetic2}
\end{align}
With dimensionless parameters $B_0 = B/\gamma$, $\kappa_0 = \kappa/k$, and $m_0 = m k /\gamma^2$, the underdamped TUR at $s=r=1$ can be written as
\begin{align}
 {\cal Q}^\textrm{u}= 	 \frac{\textrm{Var}[\Theta_\tau]}{ g^\textrm{mag}\langle \Theta_\tau \rangle^2} \left( \Delta S_\tau^\textrm{tot}  + \mathcal{I}^\textrm{mag} \right) \geq 2 ,\label{eq:magnetic_TUR}
\end{align}
with
\begin{align}
g^\textrm{mag} &=\left( \frac{1+  \kappa_0^2 m_0}{1 +  \kappa_0 B_0 -  \kappa_0^2 m_0} \right)^2,~\\ 
\mathcal{I}^\textrm{mag} &= \frac{2 \kappa_0^2 [B_0^2 - 2\kappa_0 m_0 B_0   + 2m_0 (1+\kappa_0^2 m_0 )]}{(1+ \kappa_0 B_0 -  \kappa_0^2 m_0)^2}~.
\end{align}
The derivation of $\mathcal{I}^\textrm{mag}$ is shown in Ref.~\cite{SupplMat}, and $\textrm{Var}[\Theta_\tau]$ can be also calculated for any finite $\tau$ by solving rather complex matrix differential equations numerically (not shown here, but see Ref.~\cite{Park2021Thermodynamic} for a sketch of derivations.) The EP is given by the Clausius EP with the odd-parity choice of
$B$~\cite{ChunNoh, JSLee_underdamped}, thus we obtain $\Delta S_\tau^\textrm{tot}= {\langle \Theta_\tau \rangle}/{T}$
as the average heat current is equal to the average work current in the steady state.

\begin{figure}
\includegraphics*[width=\columnwidth]{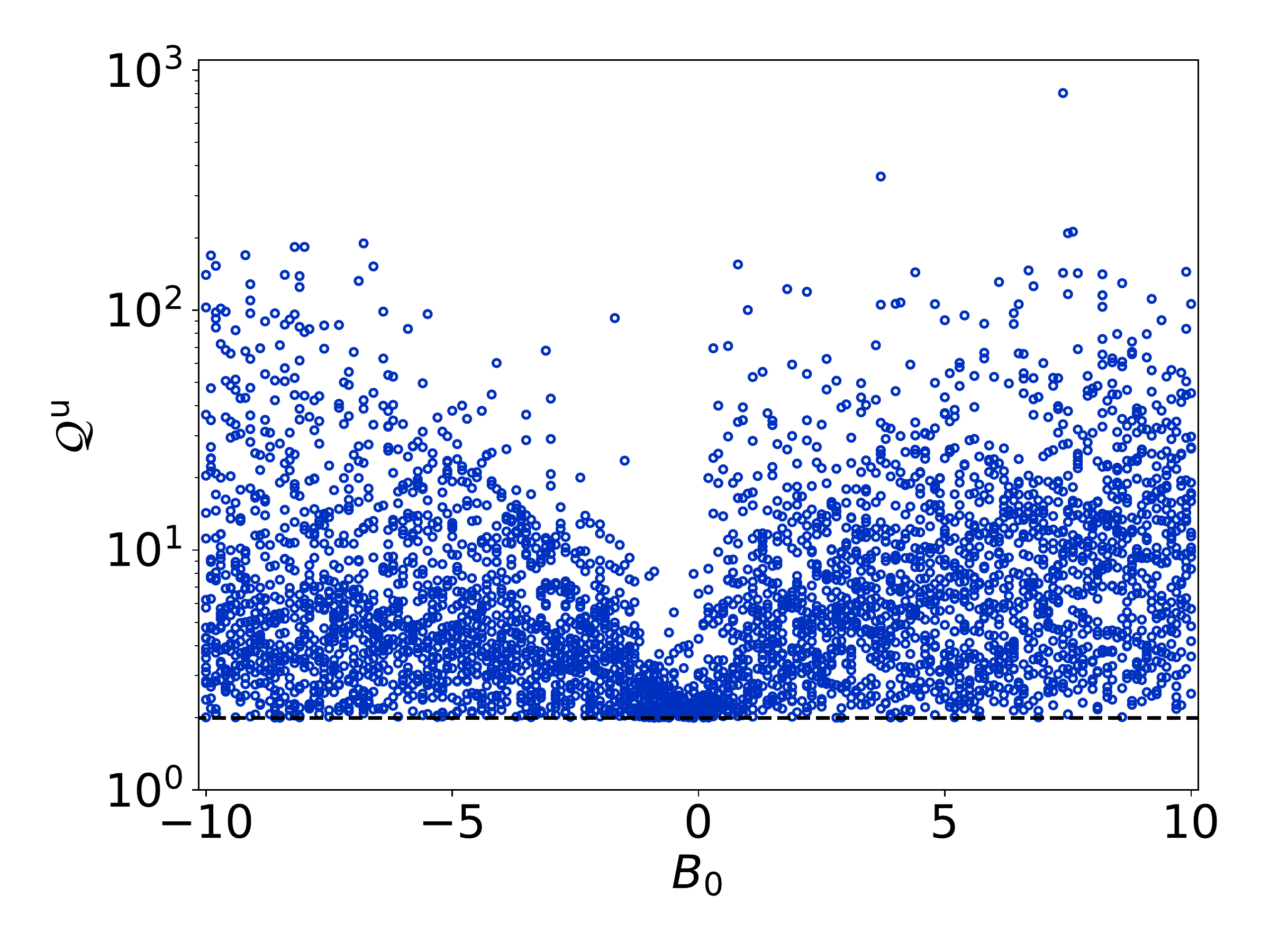}
\caption{ Plot for $Q^\textrm{u}$  of the charged particle in a magnetic field evaluated at various values of the system parameters and the observations time against $B_0$. The black dashed line indicates the lower bound of $2$.
}
\label{fig:ex2}
\end{figure}

In Fig.~\ref{fig:ex2}, we plot $\mathcal{Q}^\textrm{u}$ evaluated at various values of parameters against $B_0$.
The parameter values of $m_0$, $\kappa_0$, and $\tau$ are randomly selected from the uniform distribution with ranges of
$[0, 1]$, $[0, 10]$, and $[0, 10]$, respectively, with fixed $\gamma = k =  T = 1$. All points stay above the lower bound of $2$,
which turns out to be a very tight one for any value of $B_0$.
In the large-$\tau$ limit, $\mathcal{I}^\textrm{mag}$ is negligible and $\textrm{Var}[\Theta_\tau]$ takes a simple form~\cite{Chun}.
Then, the conventional TUR factor becomes
\begin{align}
	\mathcal{Q} = \frac{\textrm{Var}[\Theta_\tau]}{\langle \Theta_\tau \rangle^2}  \Delta S_\tau^\textrm{tot}    = 2 \frac{1+ \kappa_0^2 (1+ 3 m_0 )+  \kappa_0^3 m_0 B_0}{(1 + \kappa_0 B_0 - \kappa_0^2 m_0 )^2}~, \label{eq:magnetic_product}
\end{align}
which is larger than $2g^\textrm{mag}$ under the stability condition $1+ \kappa_0 B_0 -  \kappa_0^2 m_0 >0$, which confirms
our underdamped TUR, but can be smaller than the conventional lower bound of 2 for $\kappa_0 B_0>0$.
The previous bound including dynamical activity~\cite{Hasegawa_underdamped, JSLee_underdamped} is very loose compared to our bound here
(see Fig.S1 in Ref.~\cite{SupplMat}).
It is interesting to note that, in the equilibrium limit ($\kappa\rightarrow 0$), $g^\textrm{mag}\approx 1$ and
${\cal Q}^\textrm{u}\simeq {\cal Q}$ approaches 2 for large $\tau$.

In the zero-mass limit ($m_0=0$), we get $g^\textrm{mag}=1/(1+\kappa_0 B_0 )^2$ and
$\mathcal{I}^\textrm{mag} = 2 \kappa_0^2 B_0^2 /(1+ \kappa_0 B_0 )^2$. Thus, the original TUR
is restored when $B_0=0$ (no velocity-dependent force).
With nonzero $B_0$, the broken time-reversal symmetry due to the Lorentz force is known to
lower the TUR bound even in the overdamped limit~\cite{Chun, Park2021Thermodynamic}.
Very recently, its lower bound for the conventional TUR factor ${\cal Q}$
is rigorously obtained as $2/(1+B_0^2)$ for general nonlinear forces with a finite $\tau$~\cite{Park2021Thermodynamic}.
Our underdamped TUR also gives a lower bound for $\mathcal{Q}$ from Eq.~\eqref{eq:magnetic_TUR},
which may be tighter than the above rigorous bound for the overdamped limit, depending on the parameter values.

\subsection{Example 3: Molecular refrigerator}
We consider an one-dimensional Brownian particle driven by a velocity-dependent force $F=-\alpha v$, which serves as an effective frictional force ($\alpha>0$) to reduce thermal fluctuations of mesoscopic systems such as a suspended mirror of interferometric detectors~\cite{Cohadon, Pinard} and an atomic-force-microscope (AFM) cantilever~\cite{Mertz, Jourdan}. Thus, this mechanism is often refereed to \emph{molecular refrigerator}~\cite{Liang}.

We take $\alpha$ as an odd-parity parameter to derive a useful bound for the TUR factor~\cite{JSLee_underdamped},
which implies that the sign of $\alpha$ should change under time reversal.
Then, $F^\textrm{rev} = -s\alpha v$ and $F^\textrm{ir} = 0$ with the scale parameter $s$ for the reversible force.
The steady-state distribution is simply given by
\begin{align}
	P^\textrm{ss}(v;s) = \sqrt{\frac{m}{ 2\pi T^\textrm{e} }} \exp\left( -\frac{m}{2T^\textrm{e}} v^2 \right), \label{eq:MolRefPDF}
\end{align}
where $T^\textrm{e} = \gamma T /(\gamma + s\alpha)$ is the effective temperature.

The current of our interest is the work current done by the driving force, thus $\Lambda = -s\alpha v$,
which yields
\begin{align}
	\langle \Theta_\tau \rangle = -\tau s \alpha \langle v^2 \rangle^\textrm{ss} = -\tau s \alpha \frac{T^\textrm{e}}{m} = -\frac{\tau s \alpha \gamma T}{m(\gamma+s\alpha)}.
	 \label{eq:MolRefWork}
\end{align}
Then, we find
\begin{align}
	\Omega_\tau^\textrm{ss} = (1-s\partial_s )\langle \Theta_\tau \rangle = \frac{s \alpha}{\gamma+s \alpha} \langle \Theta_\tau  \rangle . \label{eq:Xi_MolRef}
\end{align}
By plugging Eq.~\eqref{eq:Xi_MolRef} into Eq.~\eqref{eq:main_result}, the TUR for the molecular refrigerator becomes at $s=1$
\begin{align}
	\mathcal{Q}^\textrm{u} = \frac{\textrm{Var}[\Theta_\tau]}{g^\textrm{mr}\langle \Theta_\tau  \rangle^2} \left( \Delta S_\tau^\textrm{tot}  + \mathcal{I}^\textrm{mr} \right) \geq 2,\label{eq:MolRef_TUR}
\end{align}
where $\Delta S^\textrm{tot} = \tau \alpha^2/[m(\gamma+\alpha)]$~\cite{Kim,JSLee_underdamped} and $\mathcal{I}^\textrm{mr} = \alpha^2 / (\gamma+\alpha)^2 = g^\textrm{mr}$ (see Ref.~\cite{SupplMat}). We remark that this EP is often called
the entropy pumping~\cite{Kim}. The variance $\textrm{Var}[\Theta_\tau]$ is
explicitly shown in Ref.~\cite{SupplMat}.

\begin{figure}
\includegraphics*[width=\columnwidth]{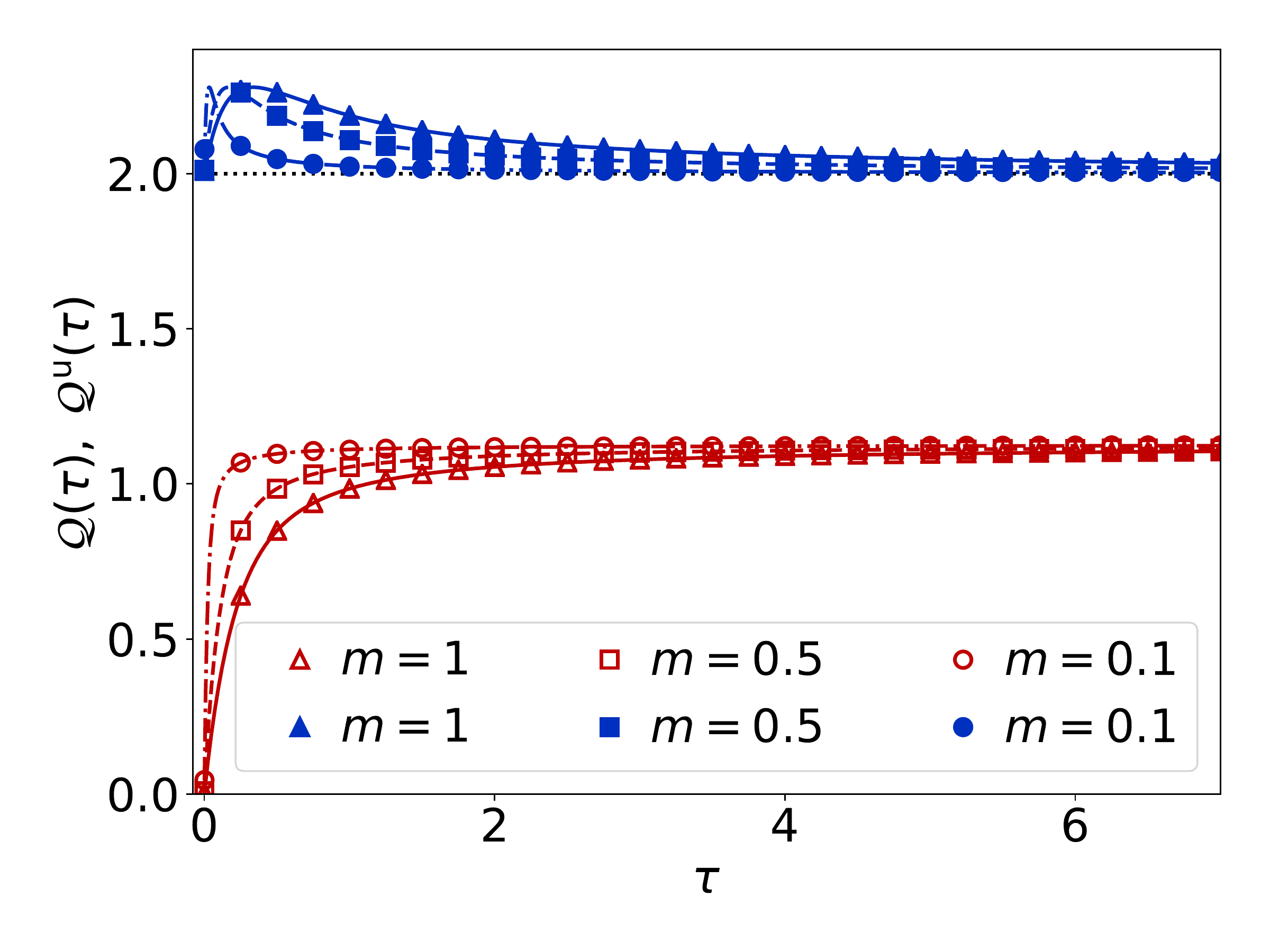}
\caption{ Plot of $\mathcal{Q}^\textrm{u}$(blue filled symbol) and $\mathcal{Q}$ (red open symbol) against $\tau$ for the molecular refrigerator. Solid, dashed, and dash-dotted curves represent analytic results for $m = 1$, $0.5$, and $0.1$, respectively.
The other parameters are set as $\alpha = 3$ and $\gamma= T = 1$.
}
\label{fig:ex3}
\end{figure}

Figure~\ref{fig:ex3} shows analytic (curves) and numerical (dots) plots of $\mathcal{Q^\textrm{u}}$ and $\mathcal{Q}$ for various values of $m$ as a function of $\tau$. The analytic results are presented in Ref.~\cite{SupplMat} and the numerical data are obtained by averaging over $10^7$ trajectories from the Langevin equation. The underdamped TUR holds for any $\tau$ as expected.
The conventional TUR factor $Q$ monotonically increases with $\tau$ and approaches $2 g^\textrm{mr}<2$.
The zero-mass limit does not lead to the original TUR due to the presence of a velocity-dependent force.

\section{Derivation of TUR}\label{sec:deriation}
The Fokker-Planck (FP) equation of the probability distribution function $P_t = P(\textit{\textbf{x}},\textit{\textbf{v}},t;s,r,\omega)$ for the  Langevin equation, Eq.~\eqref{eq:underdamped_Langevin}, can be written as
\begin{align}
	\partial_t P_t =  \mathcal{L} P_t=\sum_i(\mathcal{L}_i^\textrm{rev} + \mathcal{L}_i^\textrm{ir}) P_t, \label{eq:FPunderdamped}
\end{align}
where the FP operator  $\mathcal{L}$ is split into the reversible and irreversible parts as
\begin{align}
\mathcal{L}_i^\textrm{rev} &=   -\left[\partial_{x_i} v_i  + \left(\frac{s}{m_i}\right) \partial_{v_i} f_i^\textrm{rev} (r\textit{\textbf{x}},\textit{\textbf{v}},\omega t)\right] \\
\mathcal{L}_i^\textrm{ir} &=   -\frac{1}{m_i} \partial_{v_i} \left[ f_i^\textrm{ir} (r\textit{\textbf{x}},\textit{\textbf{v}},\omega t ) -\gamma_i v_i - \frac{\gamma_i T_i}{m_i} \partial_{v_i} \right].\label{eq:Lir}
\end{align}

Now, we consider a {\em modified} dynamics, satisfying the following FP equation parameterized by $\theta$,
\begin{align}
	\partial_t P_{t,\theta} = \sum_i\left[ \mathcal{L}_i^\textrm{rev} + (1+\theta) \mathcal{L}_i^\textrm{ir} \right] P_{t,\theta}, \label{eq:FPunderdamped_theta}	
\end{align}
which is called  the $\theta$-process.
Then, it is straightforward to show that its solution is given by
\begin{align}\label{eq:sol}
	P_{t, \theta} =P_\theta (\textit{\textbf{x}},\textit{\textbf{v}},t;s,r,\omega)= (1+\theta)^N P\left({\textit{\textbf{x}}}_\theta,\textit{\textbf{v}}, {t}_\theta; {s}_\theta, {r}_\theta, {\omega}_\theta \right)
\end{align}
with the scaled variables and parameters as
\begin{align}
&{\textit{\textbf{x}}}_\theta=(1+\theta)\textit{\textbf{x}}, ~{t}_\theta=(1+\theta)t, \nonumber\\
&{s}_\theta=\frac{s}{1+\theta}, ~{r}_\theta=\frac{r}{1+\theta}, ~{\omega}_\theta=\frac{\omega}{1+\theta}~,
\end{align}
and the normalization factor $(1+\theta)^N$.
Note from Eq.~\eqref{eq:sol} that the initial distribution $P_{0,\theta}$ at $t=0$ is $\theta$-dependent.

This modification in the FP equation is equivalent to adding an extra force $\theta\mathcal{Y}_i$ to the original process  as
\begin{align}
	F_{i,\theta} (\textit{\textbf{x}},\textit{\textbf{v}}, t) = F_i (\textit{\textbf{x}},\textit{\textbf{v}}, t) +
\theta \mathcal{Y}_i ({\textit{\textbf{x}}}_\theta,\textit{\textbf{v}}, {t}_\theta; {s}_\theta, {r}_\theta, {\omega}_\theta)
\label{eq:theta_force}
\end{align}
where $\mathcal{Y}_i =J^\textrm{ir}_i/P_t$
with the irreversible current $J^\textrm{ir}_i$ of the original process given by $J^\textrm{ir}_i=\frac{1}{m_i}[ f_i^\textrm{ir} -\gamma_i v_i - (\gamma_i T_i/m_i) \partial_{v_i} ] P_t$.
From the Onsager-Machlup theory~\cite{Onsager}, the probability of observing a trajectory $\Gamma$ in the $\theta$-process is given by
\begin{align}
	\mathcal{P}_\theta[\Gamma] = \mathcal{N} P_{0,\theta}   \prod_{i=1}^N \exp[-\mathcal{A}_{i,\theta}[\Gamma]],
\end{align}
where  $P_{0,\theta}$ is the initial-state distribution,  $\mathcal{A}_{i,\theta} [\Gamma] = \int_0^\mathcal{\tau} dt  (m_i \dot{v}_i +\gamma_i v_i - F_{i,\theta}  )^2 / (4\gamma_i T_i)$ is the action in the Ito representation, and $\mathcal{N}$ is the normalization factor which is independent of $\theta$.
By denoting $\langle \cdots \rangle_\theta = \int \mathcal{D}\Gamma \cdots \mathcal{P}_\theta [\Gamma]$ as the ensemble average over all $\Gamma$'s in the $\theta$-process, the Cram\'er-Rao inequality can be written as~\cite{Cramer1999mathematical, Rao1945information, Hasegawa2019}
\begin{align}
	\left( \partial_\theta \langle \Theta_\tau  \rangle_\theta \right)^2 \leq \textrm{Var}_\theta[\Theta_\tau ] \langle  -\partial_\theta^2 \ln \mathcal{P}_\theta \rangle_\theta, \label{eq:CR}
\end{align}
where $\textrm{Var}_\theta [\Theta_\tau ] \equiv \langle \Theta_\tau^2\rangle_\theta -\langle \Theta_\tau  \rangle_\theta^2$. The second part of the right-hand side of Eq.~\eqref{eq:CR}, usually called the Fisher information, becomes
\begin{align}
	\langle  -\partial_\theta^2 &\ln \mathcal{P}_\theta (\Gamma) \rangle_\theta = \langle  -\partial_\theta^2 \ln P_{0,\theta}  \rangle_\theta  +  \sum_{i=1}^N \left\langle   \partial_\theta^2  \mathcal{A}_{i,\theta} [\Gamma]  \right\rangle_\theta \nonumber \\
	&= \int d \textit{\textbf{x}}_0 d\textit{\textbf{v}}_0 \frac{(\partial_\theta P_{0,\theta})^2}{P_{0,\theta}} + \frac{1}{2}\sum_{i=1}^N \int_0^\tau dt  \left\langle \frac{ \left( \partial_\theta F_{i,\theta} \right)^2}{\gamma_i T_i}  \right\rangle.
\end{align}
Therefore, at $\theta=0$, we obtain
\begin{align}
	\langle  -\partial_\theta^2 \ln \mathcal{P}_\theta (\Gamma) \rangle_\theta |_{\theta =0} =  \frac{1}{2} \left( \Delta S_\tau^\textrm{tot} + \mathcal{I} \right), \label{eq:Fisher}
\end{align}
where the total EP term $\Delta S_\tau^\textrm{tot}$~\cite{risken1989fpe, Sasa_Entropic} and the initial-state dependent term $\mathcal{I}$ are given by
\begin{align}
	\Delta S_\tau^\textrm{tot} &= \sum_{i=1}^N   \int_0^\tau  dt \left\langle  \frac{ (m_i J^\textrm{ir}_i)^2} {\gamma_i T_i P_t^2 } \right\rangle,\nonumber \\
	\mathcal{I} &=  2\int d \textit{\textbf{x}}_0 d\textit{\textbf{v}}_0 \frac{(\partial_\theta P_{0,\theta})^2|_{\theta=0}}{P_0}. \label{eq:totEPandY}
\end{align}
Note that $\Delta S_\tau^\textrm{tot}$ is a time-extensive quantity while $\mathcal{I}$ is not, thus, $\mathcal{I}$ becomes negligible compared to $\Delta S_\tau^\textrm{tot}$ in the large-$\tau$ limit.

Next, we consider the average current $\langle \Theta_\tau \rangle_\theta$ in the $\theta$-process. This is a function of the scale parameters, which can be written as
\begin{align}
	&\langle \Theta_\tau \rangle_\theta (s,r,\omega) = \int_0^\tau dt \int d \textit{\textbf{x}} d\textit{\textbf{v}}~ s{\bm \chi} (r\textit{\textbf{x}},\textit{\textbf{v}},\omega t) \cdot \textit{\textbf{v}} P_{t,\theta},\nonumber \\
	 &= \int_0^{\tau_\theta} dt_\theta \int d \textit{\textbf{x}}_\theta d\textit{\textbf{v}}~ s_\theta {\bm \chi} (r_\theta \textit{\textbf{x}}_\theta,\textit{\textbf{v}},\omega_\theta t_\theta) \cdot \textit{\textbf{v}} P(\textit{\textbf{x}}_\theta,\textit{\textbf{v}}, t_\theta;s_\theta,r_\theta,\omega_\theta), \nonumber \\
	 &=\langle \Theta_{\tau_\theta}  \rangle (s_\theta,r_\theta,\omega_\theta)~.
	 \label{eq:current_theta}
\end{align}
For the second equality of Eq.~\eqref{eq:current_theta},  we take variable changes of $\textit{\textbf{x}}$ by $\textit{\textbf{x}}_\theta$ and $t$ by $t_\theta$,
and use the relations of $r\textit{\textbf{x}}=r_\theta \textit{\textbf{x}}_\theta$, $\omega t=\omega_\theta t_\theta$, and
Eq.~\eqref{eq:sol}. As $t_\theta$ and $\textit{\textbf{x}}_\theta$ are dummy variables in the integration, we get the final equality
with the average current in the original process with the scaled parameters $s_\theta$, $r_\theta$, $\omega_\theta$, and
the scaled observation time $\tau_\theta=\tau (1+\theta)$.
By differentiating the average current with respect to $\theta$ and then setting $\theta=0$, we find
\begin{align}
	\partial_\theta \langle \Theta_\tau  \rangle_\theta |_{\theta = 0} = \hat{h}_\tau  \langle \Theta_\tau \rangle =\Omega_\tau , \label{eq:diff_current}
\end{align}
where the operator $\hat{h}_\tau$ is given by $\hat{h}_\tau =   \tau \partial_\tau - s \partial_s - r\partial_r -\omega \partial_\omega$.
Using  Eq.~\eqref{eq:CR}, Eq.~\eqref{eq:Fisher}, and Eq.~\eqref{eq:diff_current}, we obtain the first main result of Eq.~\eqref{eq:main_result}.

In order to find the TUR in the overdamped limit, we consider the case where the force and the current weight function are velocity-independent (thus, $f_i^\textrm{ir}=0$) as
\begin{align}
	F_i = s f_i^\textrm{rev} (r \textit{\textbf{x}}, \omega t),~~\Lambda_i = s \chi_i (r \textit{\textbf{x}}, \omega t) \label{eq:overdamped_force}.
\end{align}
The corresponding overdamped FP equation of the probability distribution function $\rho_t=\rho(\textit{\textbf{x}},t; s,r,\omega)$ in the zero-mass limit is given as
\begin{align}
	\partial_t \rho_t =\sum_i\mathcal{L}^\textrm{o}_i \rho_t, \label{eq:FPoverdamped}
\end{align}
where the FP operator  $\mathcal{L}^\textrm{o}_i$ is given as
\begin{align}
\mathcal{L}^\textrm{o}_i =   -\frac{1}{\gamma_i}\partial_{x_i}\left[ s f_i^\textrm{rev} (r \textit{\textbf{x}}, \omega t)  -T_i \partial_{x_i} \right]~.
\end{align}

The overdamped limit of the underdamped $\theta$-process can be obtained formally by the standard small-mass expansion method using
the Brinkman's hirarchy~\cite{risken1989fpe}. In the presence of a velocity-dependent force such as a magnetic Lorentz force, the overdamped limit could become quite subtle~\cite{Park2021Thermodynamic}, which is not considered here.
However, with no irreversible force ($f_i^\textrm{ir}=0$), it can be easily seen from Eq.~\eqref{eq:FPunderdamped_theta} and
Eq.~\eqref{eq:Lir} that the $\theta$-process is simply given by the original process with the replacement of $\gamma_i$ by $(1+\theta)\gamma_i$. Thus, we can immediately write down the FP equation for the $\theta$-process in the overdamped limit as
\begin{align}
	\partial_t \rho_{t,\theta} =  \left(\frac{1}{1+\theta} \right)\mathcal{L}^\textrm{o} \rho_{t,\theta}. \label{eq:FPoverdamped_theta}
\end{align}
This is exactly the same as the \emph{virtual-perturbation} FP equation in Ref.~\cite{Seifert2020,Hasegawa2019, CR2, CR3}
with the relation of $1+\epsilon=1/(1+\theta)$ (the perturbation parameter $\epsilon$). This clearly shows that the $\theta$-process of Eq.~\eqref{eq:FPunderdamped_theta} in the underdamped dynamics is a natural extension of that in the overdamped dynamics.
This $\theta$-dynamics is simply related to the $\theta=0$ dynamics by rescaling the time $t$ by a factor of $1+\theta$. Thus, its solution
is given by
\begin{align}\label{eq:simple}
\rho_{t,\theta}=\rho_\theta(\textit{\textbf{x}},t; s, r, \omega)=\rho (\textit{\textbf{x}},\tilde{t}_\theta; s, r, \tilde{\omega}_\theta)~,
\end{align}
with the scaled parameters of $\tilde{t}_\theta=t/(1+\theta)$ and $\tilde{\omega}_\theta=(1+\theta)\omega$.
As we do not need any rescaling for $s$ and $r$, we can set $s=r=1$ from the beginning in Eq.~\eqref{eq:overdamped_force},
and the initial distribution for the $\theta$-process can be chosen to be independent of $\theta$ in general.
As also shown in Ref.~\cite{Seifert2020}, we can easily obtain
\begin{align}
	\Omega_\tau = -(\tau\partial_\tau -\omega\partial_\omega) \langle \Theta_\tau   \rangle ~\textrm{ and }~ \mathcal{I} = 0~, \label{eq:overdamped_Omega_transient}
\end{align}
which becomes Eq.~\eqref{eq:overdamped_Omega} in the steady state without any time-dependent protocol and weight function ($\omega=0$).

From the underdamped solution in Eq.~\eqref{eq:sol}, we can also find another overdamped solution of
$\rho_{t,\theta}=(1+\theta)^N \rho\left({\textit{\textbf{x}}}_\theta, {t}_\theta; {s}_\theta, {r}_\theta, {\omega}_\theta \right)$ satisfying ~Eq.~\eqref{eq:FPoverdamped_theta},
which requires the rescaling of $s$ and $r$. The initial distribution is intrinsically $\theta$-dependent
due to the dependence of ${\textit{\textbf{x}}}_\theta$, $s_\theta$, and $r_\theta$. Using this solution, we find the same formula for the TUR as in Eq.~\eqref{eq:main_result}
for the underdamped dynamics. This TUR is different from the TUR from Eq.~\eqref{eq:overdamped_Omega_transient} in general.
However, if one chooses a $\theta$-independent initial distribution, then the time evolutions of the two different solutions should be identical due to the uniqueness of the time evolution of the $\theta$-dynamics, i.e.~$\rho_{t,\theta}=\rho (\textit{\textbf{x}},\tilde{t}_\theta; s, r, \tilde{\omega}_\theta)=(1+\theta)^N \rho\left({\textit{\textbf{x}}}_\theta, {t}_\theta; {s}_\theta, {r}_\theta, {\omega}_\theta \right)$, starting from the same initial
condition. The steady-state distribution of
the $\theta$-dynamics without a time-dependent protocol is such a case, i.e.~$\rho_\theta^\textrm{ss}$ is  $\theta$-independent; $\rho_\theta^\textrm{ss} (\textit{\textbf{x}})=\rho^\textrm{ss}(\textit{\textbf{x}})$, which is obvious from Eq.~\eqref{eq:FPoverdamped_theta}.
Therefore, if a process starts from a steady state at $t=0$ and then an arbitrary time-dependent protocol is applied to the process for $t>0$, which is a usual experimental setup, both solutions become identical, leading to the same TUR in Eq.~\eqref{eq:overdamped_Omega_transient}.
Without any time-dependent protocol and weight function, this yields the original TUR in Eq.~\eqref{eq:originalTUR}. In the first two examples, we explicitly show the recovery of the original TUR in the zero-mass limit
for $\Omega_\tau^\textrm{ss}$ and $\mathcal{I}$, starting from the steady state.

\section{Conclusion}
We  derived the TUR for general underdamped Langevin systems with an arbitrary time-dependent driving from an arbitrary initial state, including velocity-dependent forces. In contrast to the previously reported one, our result is experimentally accessible and its lower bound is much tighter. Therefore, this bound can be utilized to facilitate inferring the EP by measuring a current statistics and its response to a slight change of various system parameters.
Furthermore, the original TUR for the overdamped Langevin dynamics can be understood as its
zero-mass limit. This implies that our underdamped TUR provides a universal form of the trade-off relation for general Langevin systems. It would be interesting to extend our result to systems with non-Markovian environmental noises such as active-matter systems, which are known to be described by effective underdamped Langevin dynamics~\cite{Mandal,Fodor}.

\begin{acknowledgments}
Authors acknowlege the Korea Institute for Advanced Study for providing computing resources (KIAS Center for Advanced Computation Linux Cluster System).
This research was supported by the NRF Grant No. 2017R1D1A1B06035497 (H.P.) and the KIAS individual Grants No. PG013604 (H.P.), No. PG074002 (J.M.P.), and No. PG064901 (J.S.L.) at Korea Institute for Advanced Study.

\end{acknowledgments}

\bibliography{underdamped_TUR}

\end{document}


\title{Supplemental material for ``Universal form of thermodynamic uncertainty relation for Langevin dynamics''}

\author{Jae Sung Lee$^{1}$}
\email{jslee@kias.re.kr}
\author{Jong-Min Park$^{1}$}
\author{Hyunggyu Park$^{1}$} \email{hgpark@kias.re.kr}
\affiliation{$^1$School of Physics and Quantum Universe Center, Korea Institute for Advanced Study, Seoul 02455, Korea}

\newcommand{\red}[1]{{\color{red}#1}}

\maketitle

\section*{Langevin systems driven by a force without  position dependence}

Here, we consider a case where a Brownian particle is driven by an external force without position dependence. Then, the motion can be described by the following Langevin equation:
\begin{align}
	m_i \dot{v}_i = F_i (\textit{\textbf{v}},t)-\gamma_i v_i +\xi_i, \label{eqS:Langevin}
\end{align}
where $F_i = F_i^\textrm{rev} (\textit{\textbf{v}},t) + F_i^\textrm{ir} (\textit{\textbf{v}},t)  $ with  $ F_i^\textrm{rev} (\textit{\textbf{v}},t)=  s f_i^\textrm{rev} (\textit{\textbf{v}},\omega t)$ and $ F_i^\textrm{ir} (\textit{\textbf{v}},\omega t)=  f_i^\textrm{ir} (\textit{\textbf{v}},\omega t)$.
The steady state  of such a system should be uniform in the position space, but  mathematically not well defined due to the normalization
problem without a trapping potential. To avoid this problem, one may introduce a confined space with proper boundary conditions or
consider a reduced state space spanned by velocity variables only. We take the second strategy to derive
the TUR for simplicity. Then, every derivation step is parallel to that for the general underdamped systems,
except for removing all position variables as well as the associated scale factor $r$ and the normalization constant
$(1+\theta)^N$ for the modified dynamics. Also the integration over all position variables should be also omitted.

\section*{Calculation of initial-state dependent terms and TUR factors}

\subsection*{Free diffusion with drift}

As the external force of this example is  position-independent and has a component only along the $x_1$-axis,
we consider a state space spanned by the velocity variable $v_1$ only.
First, it is easy to find the steady-state distribution $P^\textrm{ss}$ as
\begin{align}
	P^\textrm{ss} (v_1;s) = \sqrt{\frac{m_1}{2 \pi  T_1}} e^{-\frac{m_1}{2 T_1} \left(v_1 - \frac{sf}{\gamma_1} \right)^2}. \label{eqS:Pss1}
\end{align}
It is straightforward to obtain $\Delta S^\textrm{tot}_\tau=\tau s f \langle v_1 \rangle^\textrm{ss}/T_1=\tau s^2 f^2/(T_1\gamma_1)$,
$\langle \Theta_\tau \rangle  = \tau s\langle v_1 \rangle^\textrm{ss}= \tau s^2  f/\gamma_1$, and
\begin{align}
\textrm{Var}[\Theta_\tau]=2\tau s^2 \frac{T_1}{\gamma_1}\left[1-\frac{1}{\tau_0^\textrm{fr}}(1-e^{-\tau_0^\textrm{fr}})\right]\quad\textrm{with}\quad
\tau_0^\textrm{fr}=\frac{\tau\gamma_1}{m_1}~.
\end{align}
Thus, the conventional TUR factor ${\cal Q}$ is given as
\begin{align}
	\mathcal{Q} = \frac{\textrm{Var}[\Theta_\tau]}{\langle \Theta_\tau \rangle^2}  \Delta S_\tau^\textrm{tot} =2\left[1-\frac{1}{\tau_0^\textrm{fr}}(1-e^{-\tau_0^\textrm{fr}})\right]\equiv \mathcal{Q}^\textrm{fr}~,
	\label{eq:Q_free}
\end{align}
which was conjectured  in Ref.~\cite{Chun2} as  the lower bound of ${\cal Q}$ for general underdamped dynamics in one dimension,
referred as the \emph{free-diffusion bound} conjecture of $\mathcal{Q} \ge \mathcal{Q}^\textrm{fr}$.

For our underdamped TUR factor ${\cal Q}^\textrm{u}$, we need to
calculate the initial-dependent term $\mathcal{I}^\textrm{fr}$, defined in Eq.~(34) of the main text.
As the steady-state distribution in the modified $\theta$-process can be written as $P_{\theta}^\textrm{ss} = P^\textrm{ss} (v_1;s/(1+\theta))$ in Eq.~(27) of the main text, we find
\begin{align}\label{def:y_free}
	\mathcal{I}^\textrm{fr} =  2\int  d v_1 \frac{(\partial_\theta P_\theta^\textrm{ss})^2|_{\theta=0}}{P^\textrm{ss}} = \frac{2m_1 s^2 f^2}{\gamma_1^2 T_1}.
\end{align}
Thus, we finally obtain at $s=1$
\begin{align}
	\mathcal{Q}^\mathrm{u} = \frac{\textrm{Var}[\Theta_\tau]}{\langle \Theta_\tau \rangle^2} \left (\Delta S_\tau^\textrm{tot} + \mathcal{I}^\textrm{fr} \right ) =2\left(1+\frac{2}{\tau_0^\textrm{fr}}\right)\left[1-\frac{1}{\tau_0^\textrm{fr}}(1-e^{-\tau_0^\textrm{fr}})\right]~.
	\label{eqs:Qu_analytic_free}
\end{align}
One can easily show $\mathcal{Q}^\mathrm{u} \ge 2$ as expected from our underdamped TUR, and
$\mathcal{Q}^\mathrm{u} \approx 2+\tau_0^\textrm{fr}/3$ for small $\tau_0^\textrm{fr}$
and $\approx 2+2/\tau_0^\textrm{fr}$ for large $\tau_0^\textrm{fr}$.
\eqref{eq:Q_free} and \eqref{eqs:Qu_analytic_free} are plotted in Fig.~1 of the main text.
Without the knowledge of $\textrm{Var}[\Theta_\tau]$, our underdamped TUR ($\mathcal{Q}^\textrm{u}\ge 2$) predicts that
the conventional TUR factor $\mathcal{Q}$ has a lower bound as
$\mathcal{Q}\ge 2/(1+\mathcal{I}^\textrm{fr}/\Delta S_\tau^\textrm{tot})=2/(1+2/\tau_0^\textrm{fr})$, which
is obviously satisfied with $\mathcal{Q}=\mathcal{Q}^\textrm{fr}$.

\subsection*{Charged particle in a magnetic field}

\begin{figure}
	\includegraphics*[width=0.6\columnwidth]{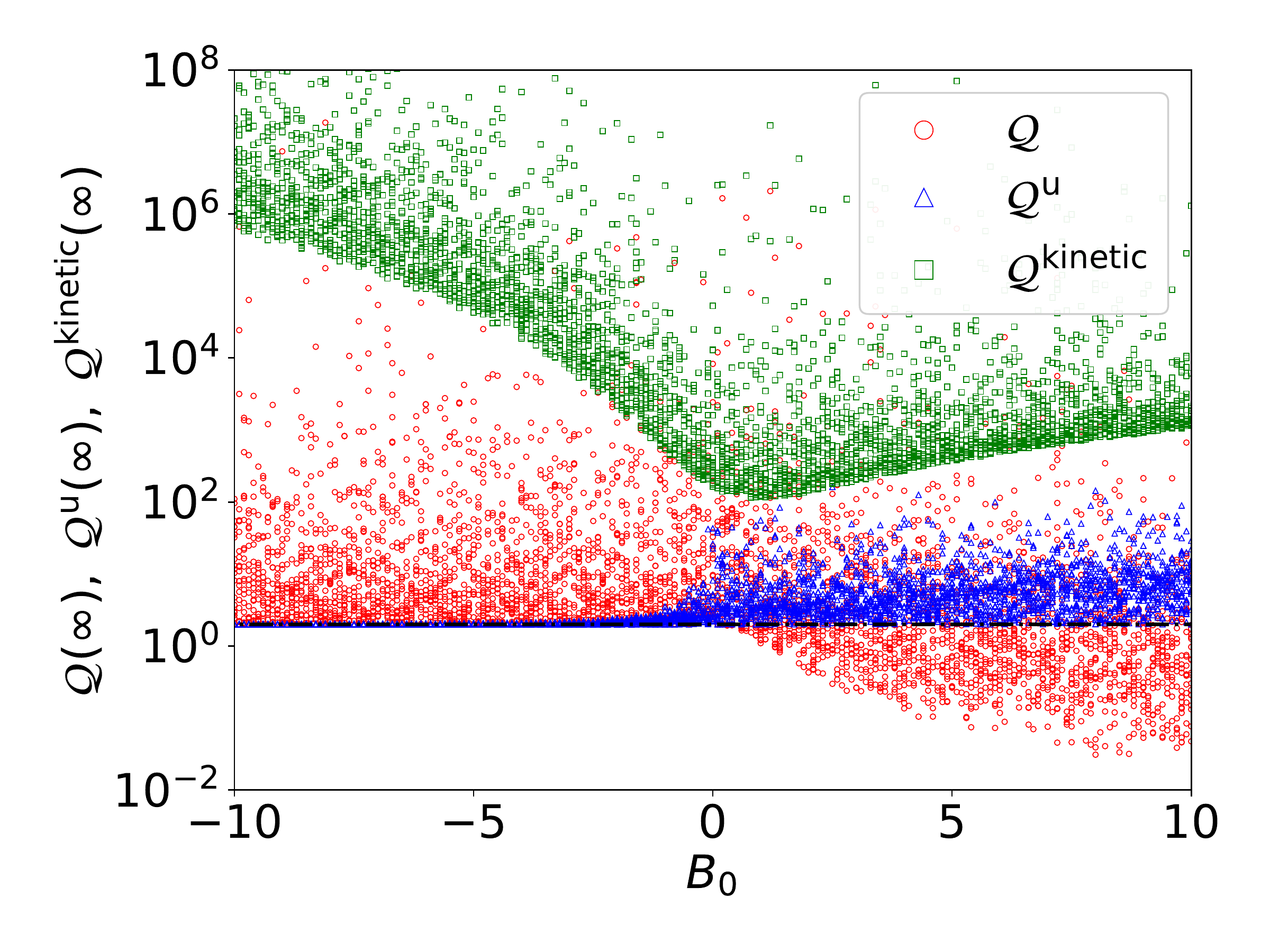}
	\caption{Plots of $\mathcal{Q}$, $\mathcal{Q^\textrm{u}}$, and $\mathcal{Q}^\textrm{kinetic}$ in the infinite-$\tau$ limit.}
	\label{fig:1SM}
\end{figure}

For a two-dimensional Brownian particle in a magnetic field, we define a state vector $\textit{\textbf{z}} \equiv (x_1, x_2, v_1, v_2)$. The covariance matrix $\textsf{C} \equiv \langle \textit{\textbf{z}}^\textsf{T} \textit{\textbf{z}}   \rangle^\textrm{ss}$ of this system is given by~\cite{Chun, JSLee_underdamped}
\begin{align}
	\textsf{C} = \frac{T}{\gamma K}
	\left( \begin{array}{cccc}
  \gamma&0& 0&-sr\kappa\\
  0&\gamma&sr\kappa&0\\
  0&sr\kappa& (\gamma sr k + s^2 r \kappa B)/m& 0\\
  -sr\kappa&0 & 0 & (\gamma sr k + s^2 r \kappa B)/m
  \end{array} \right),
\end{align}
where $K = srk+ s^2 r\kappa B/\gamma -s^2 r^2 \kappa^2 m /\gamma^2 >0$. Then, the steady-state distribution $P^\textrm{ss}$ can be written as
\begin{align}
	& P^\textrm{ss} (\textit{\textbf{z}}~;s,r)  = \frac{1}{\sqrt{\textrm{Det}(2\pi\textsf{C})}} \exp\left[-\frac{1}{2} \textit{\textbf{z}} ~\textsf{C}^{-1} \textit{\textbf{z}}^\textsf{T} \right] \nonumber \\
	&\quad = \frac{rsm (k\gamma + s \kappa B - sr m \kappa^2 /\gamma)}{4 \pi^2 T^2 \gamma}\exp \left[ -\frac{1}{2\gamma T}  \left\{ rs(x_1^2+x_2^2)(k\gamma + s B \kappa) + m(\gamma v_1^2 +\gamma v_2^2 - 2rs v_1 x_2 \kappa +2rsv_2 x_1 \kappa)  \right\}  \right].
\end{align}
The steady-state distribution for the $\theta$-process is from Eq.~(27) of the main text
\begin{align}
	P_\theta^\textrm{ss} = (1+\theta)^2 P^\textrm{ss}\left( (1+\theta)x_1,(1+\theta)x_2,v_1,v_2;\frac{s}{1+\theta}, \frac{r}{1+\theta}, \frac{\omega}{1+\theta}  \right).
\end{align}
Therefore, we find  at $s=r=1$
\begin{align}
	\mathcal{I}^\textrm{mag} = 2\int d\textit{\textbf{z}}~
	\frac{(\partial_\theta P_\theta^\textrm{ss})^2 |_{\theta =0}}{P^\textrm{ss}} = \frac{2\kappa_0^2 [B_0^2 - 2 \kappa_0 m_0 B_0  + 2m_0(1+\kappa_0^2 m_0 )]}{(1+\kappa_0 B_0 - \kappa_0^2 m_0 )^2}.
\end{align}

Figure~\ref{fig:1SM} shows the possible values of $\mathcal{Q}$ and $\mathcal{Q}^\textrm{u}$ in the $\tau=\infty$ limit for
various parameter values in the stable region ($K>0$) for $\kappa>0$. The conventional TUR factor $\mathcal{Q}$ can go below
the lower bound of 2 for $B>0$, while our underdamped TUR factor $\mathcal{Q}^\textrm{u}$ is always above the lower bound of 2
and many data are very close to the lower bound. This indicates that our bound is really tight. For comparison, we also plot
the previously found TUR factor $\mathcal{Q}^\textrm{kinetic}$ including the dynamic activity in Ref.~\cite{JSLee_underdamped} in the $\tau=\infty$ limit. As can be seen easily in Fig.~\ref{fig:1SM}, this bound is significantly looser, thus not much useful.

\subsection*{Molecular refrigerator}

Most average quantities for this system have been already reported in Ref.~\cite{JSLee_underdamped} at $s=1$ as
$\Delta S^\textrm{tot}_\tau=\tau \alpha^2/[m(\gamma+\alpha)]$, $\langle \Theta_\tau\rangle= -\tau \alpha\gamma T/[m(\gamma+\alpha)]$, and
\begin{align}
\textrm{Var}[\Theta_\tau]=2\tau  \frac{\alpha^2\gamma^2T^2}{m(\gamma+\alpha)^3}\left[1-\frac{1}{\tau_0^\textrm{mr}}(1-e^{-\tau_0^\textrm{mr}})\right]\quad\textrm{with}\quad
\tau_0^\textrm{mr}=\frac{2\tau(\gamma+\alpha)}{m}~.
\end{align}
Then, the conventional TUR factor $\mathcal{Q}$ is
\begin{align}
	\mathcal{Q} = \frac{\textrm{Var}[\Theta_\tau]}{\langle \Theta_\tau \rangle^2}  \Delta S_\tau^\textrm{tot} = 2\left ( \frac{\alpha}{\gamma+\alpha } \right )^2 \left[1-\frac{1}{\tau_0^\textrm{mr}}(1-e^{-\tau_0^\textrm{mr}})\right]~.
	\label{eqs:Q_mr}
\end{align}

The steady-state distribution for the $\theta$-process is given as
\begin{align}
	P_\theta^\textrm{ss} (v; s) = \sqrt{\frac{m}{ 2\pi T_\theta^\textrm{e} }} \exp\left( -\frac{m}{2T_\theta^\textrm{e}} v^2 \right), ~~\textrm{where } T_\theta^\textrm{e} = \frac{\gamma T }{\gamma + s\alpha/(1+\theta)}. \label{eqS:MolRefPDF}
\end{align}
Then, we find at $s=1$ 
\begin{align}
	\mathcal{I}^\textrm{mr}  =  2\int  d v \frac{(\partial_\theta P_\theta^\textrm{ss})^2|_{\theta=0}}{P^\textrm{ss}} = \left( \frac{ \alpha}{\gamma +\alpha} \right)^2~.
\end{align}
Finally, we obtain
\begin{align}
	\mathcal{Q}^\mathrm{u} = \frac{\textrm{Var}[\Theta_\tau]}{g^\textrm{mr}\langle \Theta_\tau \rangle^2} \left (\Delta S_\tau^\textrm{tot} + \mathcal{I}^\textrm{fr} \right ) =2\left(1+\frac{2}{\tau_0^\textrm{mr}}\right)\left[1-\frac{1}{\tau_0^\textrm{mr}}(1-e^{-\tau_0^\textrm{mr}})\right]~.
	\label{eqs:Qu_mr}
\end{align}
Again, it is easy to show that our underdamped TUR is always satisfied.
\eqref{eqs:Q_mr} and \eqref{eqs:Qu_mr} are plotted in Fig.~3 of the main text.
Without any information on $\textrm{Var}[\Theta_\tau]$, our underdamped TUR predicts that
$\mathcal{Q}$ has a lower bound as
$\mathcal{Q}\ge 2g^\textrm{mr}/(1+\mathcal{I}^\textrm{mr}/\Delta S_\tau^\textrm{tot})=2\alpha^2/[(\gamma+\alpha)^2(1+2/\tau_0^\textrm{mr})]$, which
is clearly satisfied with \eqref{eqs:Q_mr}. As $\alpha$ can be arbitrary, it is obvious that
the free-diffusion-bound conjecture fails in this example.

\bibliography{underdamped_TUR}